\documentclass[aps,prb,twocolumn,groupaddress,showpacs,english]{revtex4-1}

\usepackage[T1]{fontenc}
\usepackage{babel}
\usepackage{amsmath}
\usepackage{amssymb}
\usepackage{wasysym}
\usepackage{graphicx}
\usepackage{xcolor}
\usepackage{graphicx}
\usepackage{braket}
\usepackage{multirow}
\usepackage{soul}
\usepackage[normalem]{ulem}
\usepackage{enumitem}

\usepackage[linktocpage=true,
  colorlinks=true, 
  pdfborder={0 0 0},
  linkcolor=blue,
  citecolor=red,
  filecolor=yellow,
  urlcolor=blue,
  bookmarks,
  pdfauthor={},
]{hyperref}

\newcommand{\Rome}{Dipartimento di Fisica, Sapienza Universit\`a di Roma, 00185 Roma, Italy}

\begin{document}

\title{Comment on ``Simplified LCAO Method for the Periodic Potential Problem''}

\author{Shadi Qulaghasi}
\author{Giovanni B. Bachelet} \email{giovanni.bachelet@roma1.infn.it}
\affiliation{\Rome}

\date{\today}

\begin{abstract}
We report on two misprints in one of the classical, widely-used  tight-binding tables contained in the seminal, 65-years-old  paper by Slater and Koster,~\cite{Slater-Koster_PR1954} and suggest the corresponding corrections. 
\end{abstract}


\maketitle
Perhaps the simplest model of one-electron states in solids is a tight-binding hamiltonian with  a few orbitals per atom.
As early as 1929, Bloch cast the linear combination of atomic orbitals (LCAO) as an illustration of his theorem in the limit of ``strongly bound electrons'';~\cite{BlochTheorem}
25 years later, Slater and Koster
gave a key contribution to the diffusion of this method by (i) suggesting the interpretation of the
 matrix ele\-ments between atomic orbitals as adjustable parameters of a model hamiltonian,
(ii) proposing
their two-center approximation, and 
(iii) publishing their  tabulation 
for $spd$ orbitals in cubic crystals.\cite{Slater-Koster_PR1954}

We found two misprints in their Table III, which contains the hamiltonian matrix elements between Bloch sums of atomic orbitals as a function of $\bf k$ for a simple-cubic lattice. They are based
on hamiltonian matrix elements between orbitals sitting on nearest, second-nearest, and third-nearest neighboring atoms, calculated according to the two-center approximation; analogous matrix elements between Bloch sums may be immediately deduced from this table for the fcc, bcc, and diamond lattices, too.

Following the Slater-Koster notation (subscripts 1, 2, 3 for hamiltonian matrix elements involving nearest, second-nearest, and third-nearest neighbors):
\begin{itemize}[itemsep=0pt,topsep=4pt]
\item
in the third row of Table III, within the sum which defines the matrix element ($s/xy$) between two Bloch sums of $s$ and $d_{xy}$ orbitals,
the first addend should be $-2\sqrt3(sd\sigma)_2\sin\xi\sin\eta$ and not $-2\sqrt3(sp\sigma)_2$ $\sin\xi\sin\eta$, since this $\sigma$ bond derives from $s$ and $d$ orbitals and not from $p$ orbitals;
\item
in the second-to-last row of Table III, within
the sum which defines the term ($3z^2-r^2/3z^2-r^2$),
the ($dd\sigma$) parameter in one of its addends has no subscript, but it should have a 2; the correct form of the corresponding addend therefore reads $(dd\sigma)_2(\cos\xi\cos\eta + \frac{1}{4}\cos\xi\cos\zeta + \frac{1}{4}\cos\eta\cos\zeta)$.
\end{itemize}

\begin{acknowledgments}
We thank L. Boeri for useful conversations. G.B.B. acknowledges support from Fondo Ateneo-Sapienza 2017.
\end{acknowledgments}

\bibliographystyle{apsrev4-1}

\end{document}